# Electrically Modulated Thin Film Dynamics Controlling Bubble Manipulation in Microfluidic Confinement

Debapriya Chakraborty and Suman Chakraborty[*]

Department of Mechanical Engineering, Indian Institute of Technology (IIT) Kharagpur, Kharagpur 721302, INDIA

[*]Corresponding author, E-mail: suman@mech.iitkgp.ernet.in

**Abstract**

Thin film dynamics and associated instability mechanisms have triggered a wide range of scientific innovations, as attributed to their abilities of creating fascinating patterns over small scales. Here, we demonstrate a new thin film instability phenomenon governed by electro-mechanics and hydrodynamics over interfacial scales in a narrow fluidic confinement. We first bring out the essential physics of this instability mechanism, in consideration with the fact that under the action of axial electrical field in a confined microfluidic environment, perturbations may be induced on the interfaces of thin corner films formed adjacent to the walls of a microchannel, leading to the inception of ordered lateral structures. A critical electric field exists beyond which these structures from the walls of the confinement intermingle to evolve into localized gas pockets in the form of bubbles. These bubbles do not remain static with further changes in electric field, but undergo a sequence of elongation-deformation-breakup episode in a dynamically evolving manner. By elucidating the complex interplay of electro-hydrodynmic forces and surface tension, we offer further insights into a new paradigm of interfacial instability mediated controlled microbubble manipulation for on-chip applications, bearing far-ranging scientific and technological consequences in executing designed fluidic operations in confined miniaturized environment.

There has been a long interest in studying various physical facets of thin liquid films, motivated by their spontaneous formations in nature and the roles played by them in a wide variety of transport processes in practical scenarios [1]. In practice, thin films are often characterized by several special characteristics, including the formation of regular structures [2] (self-assembly), formation of shock-fronts [3] and periodic waves, and instabilities associated with the complex interplay of interfacial forces, leading to non-trivial dynamics over small scales. Applications of such dynamical aspects of thin liquid films [1] range from complex coating flows [4], optoelectronic display materials, engineering applications like heat exchangers, microfluidics [5, 6] and micro-electro-mechanical devices, gravity currents [1], granular and debris flows [3], snow avalanches [8], biophysical aspects like lung airways [9], tear-film flows, and bio-adhesion [10], to name a few. Thin films often act as precursors to create templates for patterned arrays of micro- and nano-structures. These patterns are obtained by exploiting the physical and energetic properties of the film. Alteration of these properties in tandem with other hydrodynamic effects results in instabilities (spontaneous or induced) in these films. Nevertheless, thicknesses of these films also play crucial roles in determining the concerned instability phenomena [11,12]. In practice, thin film instabilities may be triggered by various mechanisms (for example, see [13-42]), including those of non-electrical as well as electrical origin.

Here, we first bring out a new thin film instability phenomenon triggered by complex electro-mechanics and interfacial hydrodynamics in a narrow fluidic confinement. In sharp contrast to previous reports on thin film dynamics triggered by externally applied transverse electric field alone, here, we show that the interfacial thin films could be altered through an intricate interplay of externally applied *axial electric field*, spontaneously induced transverse electric field, and hydrodynamics over interfacial scales, in a confined microfluidic environment. We term the underlying phenomenon as Electro-Osmo-Hydro-Dynamic (EOHD) instability. In particular, we have discovered two new regimes of electrokinetically modulated thin film instability in a micro-confinement, in which perturbations may be induced on the interfaces of the corner films adjacent to the channel walls under the action of an axial electric field, so as to give rise to ordered lateral structures. In one of these regimes (Regime I), the volumetric electroosmotic forces in the thin film dominate over the force generated from the transverse gradient of electrical pressure (corresponding to electric field stored in the dielectric medium), whereas the relative dominance of these two opposing forces reverses in the other regime (Regime II). There exists a critical electric field demarcating these two regimes, beyond which the corrugated thin films progress from opposite walls of the microchannel and eventually inter-mingle, leading to the formation of gas pockets or bubbles. By exploiting the underlying instability mechanism dictating this phenomenon, we demonstrate the dynamics of controlled microbubble generation and manipulation, as triggered by a complex interplay of electro-hydrodynamic forces and interfacial tension. We show that these features are accompanied by a sequence of downstream events such as elongation-deformation-breakup in a dynamically evolving environment. Our results bear immense consequence towards designed microbubble manipulation for on-chip applications, which may be judiciously programmed for executing precise fluidic operations in a confined miniaturized environment.

**Results**

**Experiment.** We have considered straight rectangular glass microchannels of different channel dimensions (width $w$ and depth $H$ = 12 μm) - 50μm×10μm, 25μm×12μm, 50μm×25μm and 65μm×30μm. Fabrication of such microchannels is common and may be found in several studies related to microfluidics [43, 44]. A meniscus is formed (see Fig. 1a) by partially filling the glass microchannel with borate buffer (10 mM). For microchannel surfaces which are wetting in nature, the liquid meniscus extends in the form of precursor film over the channel walls, leaving a gas filled region even prior to the application of the electric field. A thin film, typically known as corner film (with thickness of order of 1 μm), is formed at the gutters (or corners) of the microchannel. Micropipette tips are cut and used as inlet and outlet reservoirs filled with the buffer solution, in order to complete the electrical contacts. An electric potential is applied across these reservoirs by dipping platinum wire into the solution connected to a constant voltage sourcemeter (Keithley, USA). The thin film profiles are observed in phase-contrast microscope (Olympus IX71, USA). For visualisation of the flow field, 500 nm fluorescent particles (Sigma Aldrich, India) are introduced into the buffer solution with 200 times dilution and observed with an exposure time of 500 ms.

**Thin Film Instability.** Under the action of an *axial electrical field (E),* interfaces of thin corner films get perturbed, leading to the inception of lateral structures having corrugated topographies on highly wettable substrates (see Fig. 1b). This phenomenon may be perceived as a consequence of EOHD instability mechanism, triggered by interplay between the electric field forces (on free charges or field stored in the dielectric), thin film hydrodynamic forces and the interfacial tension. The corrugated interfacial layers, formed at the opposite walls, tend to grow progressively with increments in the applied electrical voltage. Beyond a critical electric field, $E_{cr}$, the progressively thickening interfacial films eventually intermingle at the channel-centerline at an instant known as rupture time $t_r$. This rupture time is a hallmark of a transformation of the thin film (Regime I) to the meniscus formation (Regime II) (see Fig. 1c), which consequently results in localized gas pockets that evolve dynamically. The gas pockets, thus formed, subsequently reshape and form bubbles inside microchannels (see Fig. 1d). The parameters, $t_r$ and $E_{cr}$ are not constants, but are inter-dependent and are functions of the channel dimensions, as demonstrated in Fig. 2. It is observed that the rupture-time ($t_r$) decreases rapidly with the increasing electric field and empirically fitted to be as $E^{-4.3}$.

**Critical Electric Field.** We analyse the instability wavelength ($\lambda$) corresponding to the dominant spatial frequency by transforming the thin film profiles (Fig. 3) into Fourier space. For axial electric field $E < E_{cr}$, spatiotemporal evolutions of the film thickness reveal that the perturbations recur in space (without intermingling), while for $E > E_{cr}$, the thin films from the opposite walls intermingle with rupturing of the film at $t=t_r$. The instability wavelength, which is a function of the electric field $E$, is shown in Fig. 3. It is observed that the wavelength varies as $\lambda \sim E^{-1/2}$ for $E < E_{cr}$ (Regime I), while $\lambda \sim E^{-1}$ for $E > E_{cr}$ (Regime II), which indicates dominance of different forces governing the physics of instability in different regimes.

**Bubble Deformation and Breakup.** Bubble formed after intermingling of the interfaces tends to adjust to the incipient stresses by reconfiguring itself into a peanut shape. During this process, electrical forces deform each bubble while surface tension forces resist the deformation. The deformation of the neck of each bubble, $\varsigma$ (normalized by $w$), is a function of the ratio of the electrical forces to that of the surface tension forces, which may be expressed in terms of the electrical capillary number $\chi = \dfrac{\varepsilon_0 \varepsilon_1 E^2 H}{\sigma}$ where $H$ is the depth of the microchannel. It is found to scale linearly with $\chi$. Further, the deformation, $\varsigma$, is also found to scale with $\left(L_b/L_b^0\right)^{-n}$, where $L_b$ and $L_b^0$ are the length of the deformed and un-deformed bubble respectively, and the exponent $n$ may be obtained as $n = 2.1$, where all the experimental data collapse to a single master curve (see Fig. 4a). Interestingly, the bubble elongation-deformation described as above is by no means an eternal process. As such, the elongation of each bubble increases progressively till the neck deforms and becomes zero, resulting in a bubble breakup. The numbers of new pockets formed are essentially dictated by the dominant wavelength ($\lambda$) of the spatial disturbance of the thin film. The sizes of the daughter pockets (microbubbles thus generated) further scale linearly with the dominant wavelength, as evident from Fig. 4b, which implies that EOHD instability is responsible for the bubble break-up.

**Origin of the Instability.** The driving force for the instability triggering the above sequence of events is the electric field induced stress in the corner films in microchannels. The electrostatics across the solid-liquid interface (where the thin corner-films come in contact with the glass surfaces) is faceted by the fact that the glass surface develops free surface charges because of ion adsorption or reaction in presence of the ionic liquid. This surface charging, in turn, leads to a charge density distribution in the solution, resulting in the development of Electric Double Layer (EDL) [45]. Distribution of free charges within the EDL, governing the thin film dynamics, is essentially dictated by the Poisson-Boltzmann (PB) equation [45] as $\nabla^2 \psi = \kappa^2 \psi$ ($\psi$ being the potential induced in the EDL and $\kappa$ being the inverse of the Debye length describing a characteristic thickness of the EDL), consistent with the Debye-Hückel approximation [46]. Assuming the dielectric properties remain unaltered, the electrical stress may be expressed as [47]: $\mathbf{T}^E = -\Pi \mathbf{I} + \varepsilon_0 \varepsilon_1 \left( \mathbf{EE} - \frac{1}{2} E^2 \mathbf{I} \right)$ where, $\Pi$ is the osmotic pressure, $\mathbf{E}$ is the resultant electric field, $\mathbf{I}$ is the identity tensor, and $\varepsilon_1$ and $\varepsilon_0$ are the permittivities of the medium and free space respectively. The electrical stress has contribution from two effects - osmosis due to the presence of free charges in the EDL and the electrical field stored in the liquid dielectric medium. The ratio of the stress induced by the osmotic effects to that of the stress due to the electric field in the dielectric is proportional to $E^{-2}$. Further, the electric field varies inversely with the thin film thickness (for constant electric flux) and directly proportional to the applied external potential. Hence, as the film thickness decreases and/or the external potential increases, the electric field increases, and the stress experiences a shift of dominance from osmotically dominated regime to a regime dictated by the electric field established in the dielectric.

**Theoretical Modelling.** We provide a simple yet insightful theoretical analysis on the dynamical aspects of the thin corner films formed in the wall-adjacent interfacial layer. Solution of the PB equation gives the free charge distribution within the EDL. This free charge distribution influences the dynamics of the thin film in a rather profound manner, consistent with the mechanism of electroosmosis. In electroosmosis, interaction between an induced transverse field in the EDL and applied axial electrical field results in shear gradients in the wall-adjacent layer, pulling the liquid axially. The effects of thin film under electroosmosis may be quantitatively captured by considering the divergence of an electrical stress tensor ($\mathbf{T}^E$) in the momentum equation. The osmotic pressure may be calculated as: $\Pi = (n^+ + n^-) K_B T$; $n^+$ and $n^-$ being the number densities of the positive and negative ionic species, respectively, as described by the Boltzmann distribution [45]. The liquid pressure, $P_l$, alternatively interpreted as the effective pressure, is a combined consequence of the gas pressure, Laplace pressure (surface tension effect), osmotic pressure, and electrical pressure, expressed as: $P_l = P_g - \sigma h_{xx} + \Pi + p_{el}$ where the subscript $xx$ represents the double derivative of the local film thickness $h$ with respect to the axial coordinate $x$. Here, $P_g$ is the gas pressure, $p_{el} \left( = \frac{\varepsilon_0}{2} (\varepsilon_1 - 1)(E_t^2 + \varepsilon_1 E_n^2) \right)$ denotes the pressure induced from the jump in the electrical stresses at the liquid-gas interface, $\sigma$ is the liquid-gas surface tension coefficient, and $E_n$ and $E_t$ are the normal and tangential components, respectively, of the external electric field. While arriving at this expression, we also consider that at the liquid-gas interface, the electric field experiences the continuity of the displacement field $(\sigma_{el} E_n)_l = (\sigma_{el} E_n)_g$, where $\sigma_{el}$ is the electrical conductivity. For a perfect dielectric, it is consistent to replace $\frac{\sigma_{el,l}}{\sigma_{el,g}}$ by $\frac{\varepsilon_1}{\varepsilon_0}$ [48]. Further, the tangential component of the electric field is continuous across the interface: $(E_t)_l = (E_t)_g$. In the lubrication limit, the film thickness varies slowly, and the external electric field in the thin film region may be considered to primarily act along the tangential direction (i.e., normal component of the electric field, $E_n = 0$) at the solid-liquid and liquid-gas boundaries of the thin film. In our experiments, we have inserted 500 nm fluorescent particles in the flow-field to observe the local streaklines with an exposure time of 500 ms. As the streaks are all oriented in a direction tangential to the interface (see inset of Fig. 3), we assume that the interfacial electrical field, acts solely in the tangential direction, however, varies axially, owing to the fact that the local film thickness also varies axially, under the constraint of a constant electric flux ($K$), so that one may write $E_t = K/h$; $h$ being the local film thickness.

The hydrodynamics in the thin film in steady state may be described by using the equation of momentum conservation as: $-\nabla p + \nabla \cdot \mathbf{T} = 0$ (where $p$ is an effective pressure which includes the effects of interfacial tension, osmotic pressure, and electrical stresses), along with the boundary conditions - (i) $\mathbf{t} \cdot \mathbf{T} \cdot \mathbf{n} = 0$ at $y=h$ (where $\mathbf{T}$ is the resultant stress tensor, considering both hydrodynamic ($\mathbf{T^H}$) and electrical stress ($\mathbf{T^E}$), $y$ being a transverse coordinate axis with origin at one of the microchannel walls) and (ii) the velocity vector $\vec{u} = 0$ at walls ($y=0$). The above, in conjunction with the continuity equation, yields the governing equation for the 2D evolution of the corner film as:

$$\frac{\partial h}{\partial t} = \frac{\partial}{\partial x}\left(\frac{h^3}{3\mu}\frac{\partial}{\partial x}\left(-\sigma h_{xx} + p_{el}\right)\right) + U_e h_x \Theta \tag{1}$$

where $\mu$ is the dynamic viscosity of the liquid, $U_e$ represents a characteristic electroosmotic velocity scale, given by: $U_e = -\varepsilon_0 \varepsilon_1 \psi_0 (\kappa K) \tanh(\alpha)/(\alpha \mu)$, with $\alpha = e\psi_0/4K_B T$ (here $\psi_0$ is the surface potential) and $\Theta = 1 - e^{-\kappa h}(1 + \kappa h/2)$.

**Linear Stability Analysis and Dispersion Relation.** The parameter $h$, as appearing in Eq. 1, increase in response to the application of a progressively intensifying electric field, owing to the higher flux of ions in the thin film region, till the opposite films intermingle. We use a linear stability analysis to predict different instability regimes during dynamical evolution of the film as governed by Eq. 1. The base state solution $h=h_0$ satisfies the governing equation, and we check the stability of this base state with respect to small perturbations. We expand the base solution in normal modes with an ansatz: $h = h_0 + \varepsilon e^{ikx+\omega t}$ where $h_0$, $k$, $\omega$ and $\varepsilon$ are the mean film thickness, spatial growth rate, the temporal growth coefficient and the amplitude of the initial disturbance, respectively, with the assumption that $\varepsilon \ll h_0$. The dispersion relation may be obtained correct to the first order of $\varepsilon$ as: $\omega = -h_0^3 k^2 \left(\partial p_{el}/\partial h\big|_{h=h_o} + k^2 \sigma\right)/3\mu$ and $\omega = -3\mu U_e/h_o^2 - \partial p_{el}/\partial h\big|_{h=h_o}$. In the limit of $U_e \to 0$, the inverse wavelength ($k$) is a free parameter (similar to the behaviour described in Refs. [21, 24]), because the imaginary part solution is the trivial solution of the dispersion relation $\partial p_{el}/\partial h\big|_{h=h_o} + k^2 \sigma = 0$. For such a case, the dominant wavelength and maximum growth factor corresponding to the fastest growing linear mode ($\partial \omega/\partial k = 0$) may be obtained as $\lambda_{\max} = 2\pi/k = 2\pi\left(-\partial p_{el}/\partial h\big|_{h=h_o}/2\sigma\right)^{-1/2}$ and $\omega_{\max} = h_0^3 \partial p_{el}/\partial h\big|_{h=h_o}^2/12\mu\sigma$.

**Discussion.** To gain insight into the dominance of different force, we assess the relative importance of the electroosmotic term $-3\mu U_e/h_o^2$ over the electrical pressure gradient term $-\partial p_{el}/\partial h\big|_{h=h_o}$ in the dispersion relation. The order of the electrical pressure gradient may be approximated as $-\partial p_{el}/\partial h\big|_{h=h_o} \sim \varepsilon_0\varepsilon_1 E^2/h_0$. In order to have the electroosmotic effects dominant over the gradient of electrical pressure, i.e., $O(\mu U_e/h_o^2) \gg O(\partial p_{el}/\partial h\big|_{h=h_o})$, it is thus essential to have $|E| \ll |\psi_0 \kappa \tanh(\alpha)/\alpha|$. With the typical values $|\psi_0|$ = 75mV, $\kappa^{-1}$ = 100 nm, $T$=298 K, it may be found that the wavelength of instability is electroosmotically dominated (Regime I), when the magnitude of the electric field is less than $|E| \sim O(10^4)$ V/m, whereas the effects of the gradient of pressure from the jump in the electrical stresses at the liquid-gas interface ($p_{el}$) becomes significant beyond this threshold limit of the electric field (Regime II). When $O(\mu U_e/h_o^2) \ll O(\partial p_{el}/\partial h\big|_{h=h_o})$, the system essentially behaves like a conductive fluid. Thus, below the threshold electric field, the wavelength $\lambda$ and the time scale $\omega^{-1}$ are dominated by the electrosomotic effects and scale as $\lambda \sim E^{-1/2}$ and $\omega^{-1} \sim E^{-3}$ respectively (in Regime I). On the other hand, above the critical electric field, the effect of $p_{el}$ becomes dominant, resulting in $\lambda \sim E^{-1}$ and $\omega^{-1} \sim E^{-4}$ (in Regime II). These scaling expressions agree closely with the corresponding experimental predictions (as evident from Fig. 2 for $t_r$ and Fig. 3 for $\lambda$). The above timescale $\omega^{-1}$ also corresponds to the time scale of the rupturing the thin films, $t_r$ as mentioned earlier. This rupture time scale attributes a transition from the thin film regime to the meniscus formation regime, and results in the formation of localised gas-liquid pockets in the form of bubbles of different controllable dimensions (as influenced by the wavelengths of the instability mode corresponding to the electric field being axially applied).

In summary, we have utilised electroosmotic flow to induce interfacial instabilities of the corner films in microfluidic channels, triggering non-trivial morphological dynamics. We have demonstrated that the physics of the interfaces, under these circumstances, can be elegantly controlled by the application of the *axial electric field*. The consequent electro-mechanical facets of the thin film instability may essentially lead to the events of microbubble generation, elongation-deformation and break-up phenomena, which may be exploited towards realizing a new paradigm of integrated and controlled microbubble generation and manipulation for on-chip applications, with axial electric field as a tuning parameter. It is, however, also important to note that for partially-wetting or non-wetting substrates (unlike the cases of completely wetting substrates considered here), additional facets of thin film instabilities may feature, because of complicated interaction between corner film and surface film (~100 nm) instabilities. The dynamics of the instability and the consequent microbubble manipulations demonstrated in our work hold a potential for wide range of applications such as algorithmic pattern formation, programmed generation of emulsion and foam, enhanced oil recovery (by increasing mobility using microbubbles), carrying chemical payload for targeted drug delivery, and different lab-on-a-chip fluidic operations (using microbubble) like valving, dispensing and trafficking.

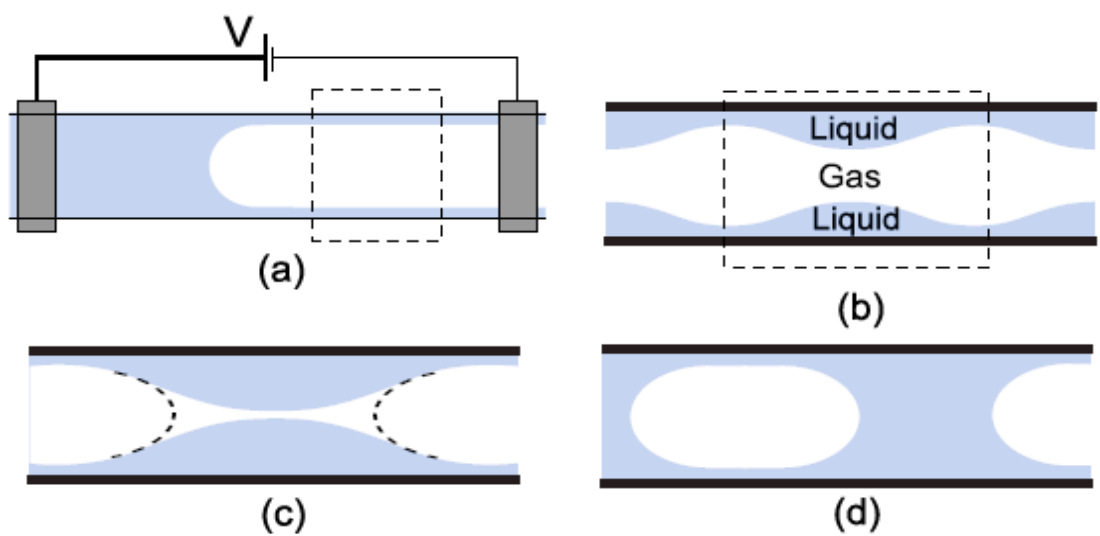

**Fig. 1.** Schematic representation of the different stages of the EOHD instability phenomenon (a) Microchannel filling using an axially applied electric field (b) The unstable corner film ahead of the meniscus (c) Intermingling of the unstable films from the top and bottom resulting in the transformation of the thin film to meniscus (d) Formation of gas pockets and reshaping into the bubbles (see Supplementary Information for experimental images)

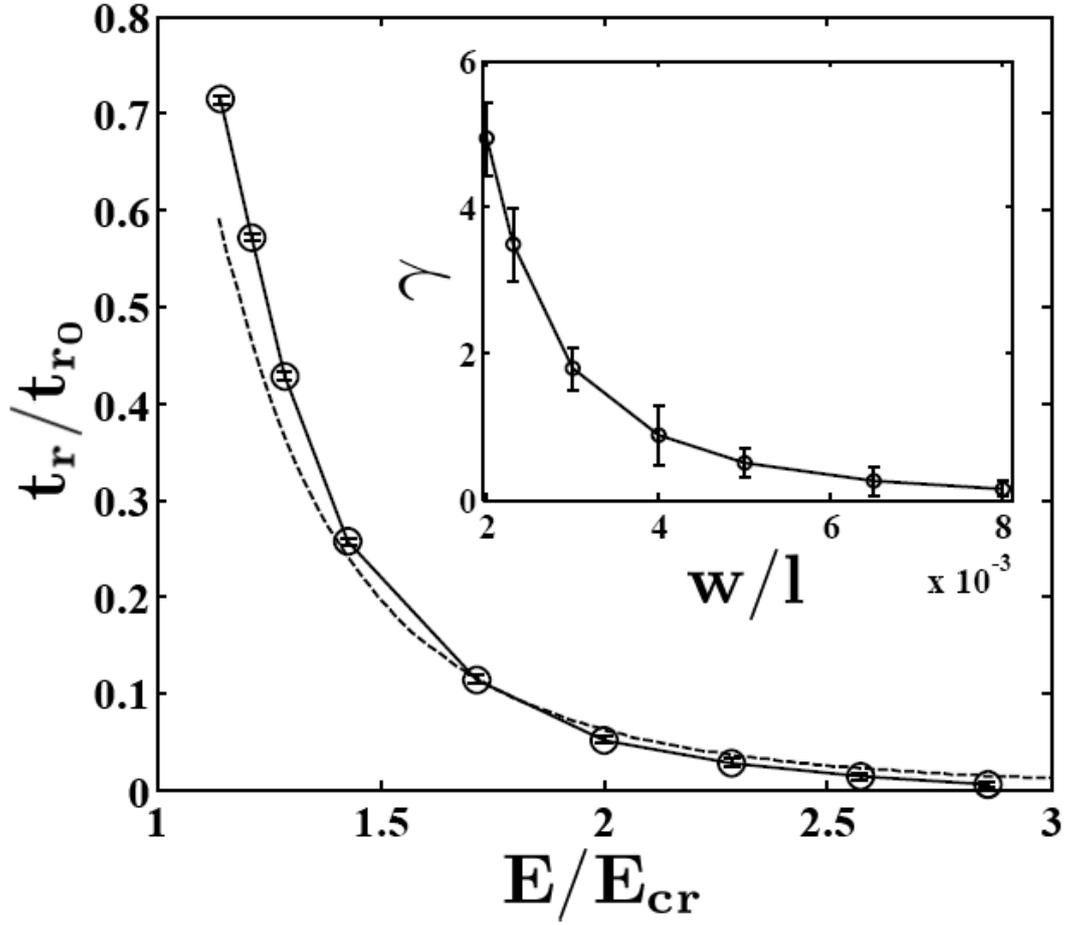

Fig. 2. Variations in rupturing time of the thin film (normalized by the rupturing time corresponding to $E=E_{cr}$, denoted by $t_{r0}$) with $E/E_{cr}$, for $E>E_{cr}$. The solid line corresponds to the empirical fit of $E^{-4.3}$ and dashed line represents theoretically predicted $E^{-4}$. Inset shows the variation of $E_{cr}$, expressed in a dimensionless form $\gamma = \dfrac{\varepsilon_m \varepsilon_0 E_{cr}^2 l^2}{\sigma H}$, as a function of the dimensionless channel width $w/l$, $H$ being the microchannel depth

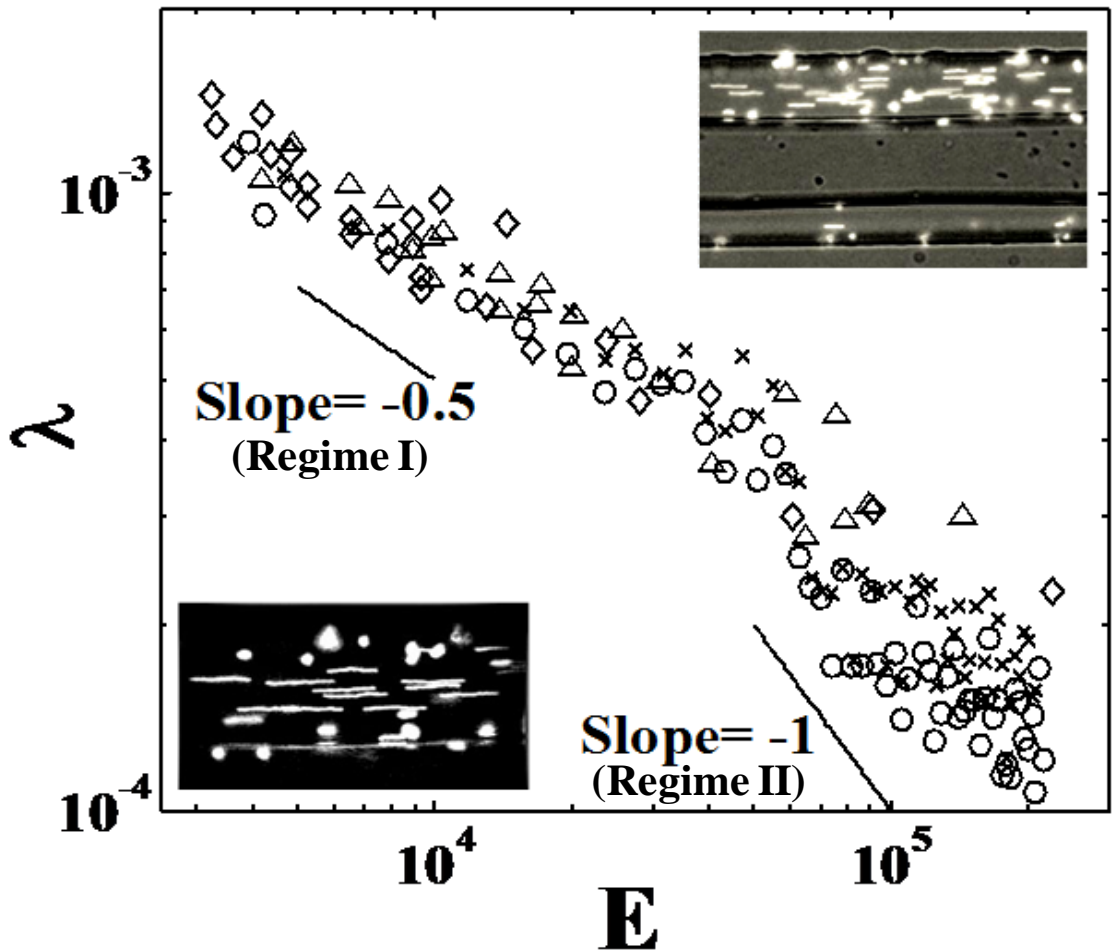

**Fig. 3.** The variation of instability wavelength (in m) as a function of electric field (in V/m). Markers correspond to the experimentally obtained data points for different channel dimensions - 50μm×10μm (○), 25μm×12μm (□), 50μm×25μm (◊) and 65μm×30μm (x) and experimental errors are within 2.5 % (errorbars not shown for clarity). The slopes of different regimes - $E < E_{cr}$ (regime I) and $E > E_{cr}$ (regime II) are shown. The top-right inset shows the bright-field image of an unstable liquid-gas-liquid interface and bottom-left inset shows the zoomed fluorescent image of the streaks of the motion of 500 nm fluorescent beads with exposure of 500 ms (to estimate the velocity and trajectory of these beads).

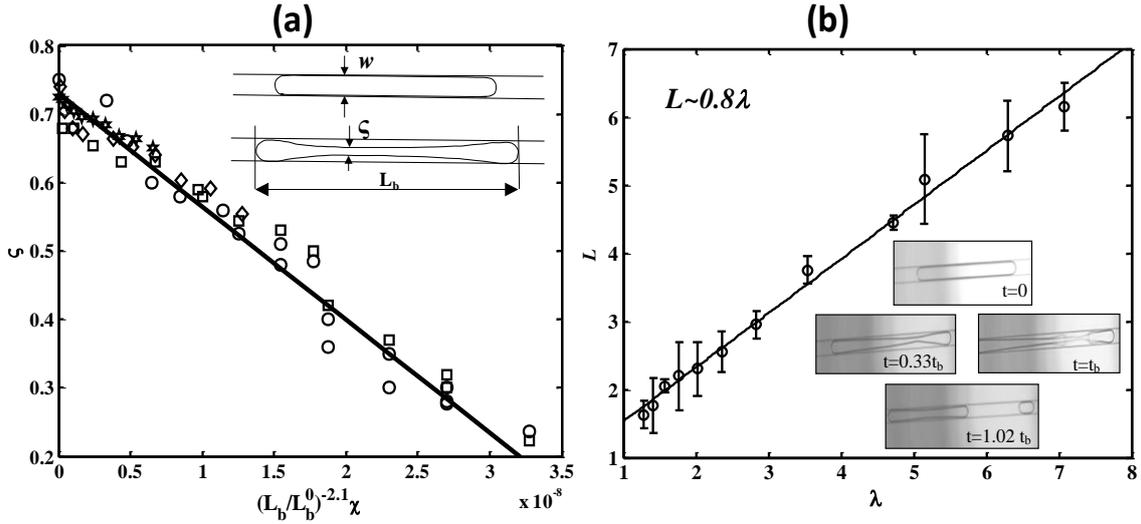

**Fig. 4** (a) Deformation of the neck ($\varsigma$) of a gas pocket formed (nondimensionalised by the width of the channel $w$) in presence of an electric field as a function of $\left(L_b/L_b^0\right)^{-2.1}\chi$, after the electric field initially elevated beyond $E_{cr}$ to trigger bubble formation, and then lowered below $E_{cr}$, for different channel dimensions - 50μm×10μm (○), 25μm×12μm (□), 50μm×25μm (◇) and 65μm×30μm (✶). Experimental errors are within 3.2% (errorbars not shown for clarity). See Supplimentary Movie for the deformation of the bubble under external electric field. The inset shows the optical micrograph of a bubble and its deformation with the application of electric field, for one representative scenario. (b) The lengths of daughter bubbles generated after break-up ($L$), as a function of $\lambda$; both normalized by $w$ (markers are not specific to any channel dimensions). The inset shows a sequence of images illustrating bubble break-up as a function of time of break-up ($t=t_b$).